\documentclass{article}
\usepackage{url}
\usepackage{amsmath}
\usepackage{graphicx}
\begin{document}
\vspace*{-3cm}
\begin{flushright}
\today

\end{flushright}


\vspace{5.0ex}

\begin{center}
  
\begin{Large}
  {\bf BAYES AND FREQUENTISM: A PARTICLE PHYSICIST'S PERSPECTIVE}
  \end{Large} \footnote{To appear in Contemporary Physics}

\end{center}

\begin{center}

\vspace{10.0ex}

\vspace{1.0ex} {\Large Louis Lyons}\\

\emph{Blackett Lab., Imperial College, London SW7 2BW, UK} \\ 
and\\
\emph{Particle Physics, Oxford OX1 3RH} \\

\vspace{0.3cm}
e-mail: l.lyons@physics.ox.ac.uk
\vspace{1.0ex} 
\end{center}

\vspace{10.0ex}

  
  


\begin{abstract}

In almost every scientific field, an experiment involves collecting data
and then analysing it. The analysis stage will often consist in trying to
extract some physical parameter and estimating its uncertainty; this is 
known as Parameter Determination. An example would be the
determination of the mass of the top quark, from data collected from
high energy proton-proton collisions. A different aim is to choose 
between two possible hypotheses. For example, are data on the recession 
speed  $s$ of distant galaxies proportional to their distance $d$, or do they fit
better to a model where the expansion of the Universe is accelerating?

There are two fundamental approaches to such statistical analyses - 
Bayesian and Frequentist. This article discusses the way they differ
in their approach to probability, and then goes on to consider how this affects
the way they deal with Parameter Determination and Hypothesis Testing. 
The examples are taken from every-day life and from Particle Physics.

\end{abstract}

\section{INTRODUCTION}

\begin{figure}[h]
\begin{center}
\includegraphics[width=1.1\textwidth]{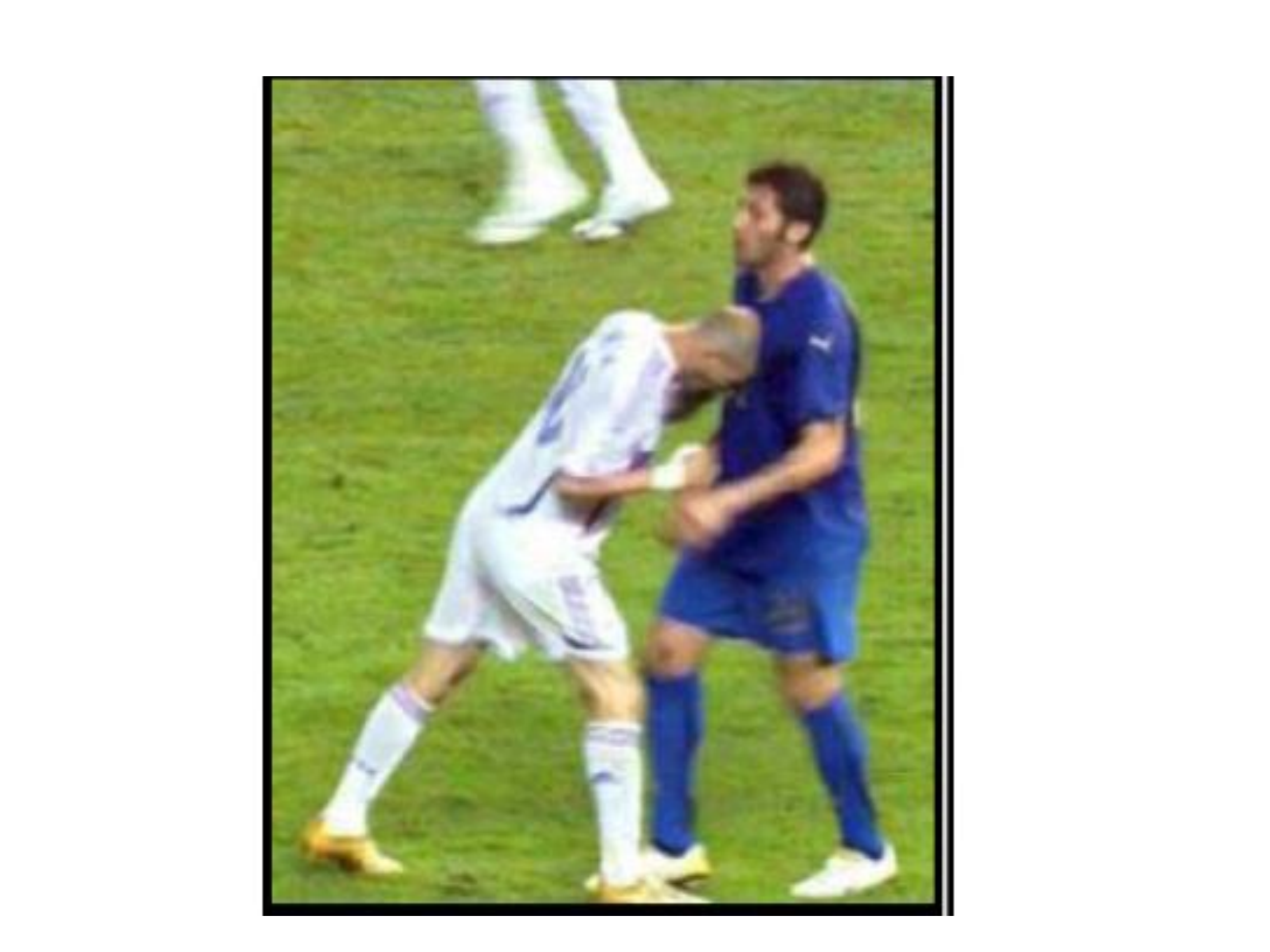}
\caption{This incident from the Wold Cup soccer final in 2006  was used to illustrate the `discussions' 
that took place between sub-groups within a
particular experiment about the relative merits of  Bayesian and Frequentist analyses\cite{World_Cup}.
\label{fig:World_Cup}    }
\end{center}
\end{figure}

There are two fundamental approaches to statistical analysis, Bayesianism
and Frequentism. The Bayesian approach dates back to
Reverend Thomas Bayes, whose paper was publishes posthumously in 1763. The
Polish statistician Jerzey Neyman played a crucial role in the 
development of frequentist statistics. 
In the past there have been vigorous discussions about 
the relative merits of these two methods (see fig. \ref{fig:World_Cup}).


\begin{figure}[h]
\begin{center}
\includegraphics[width=1.1\textwidth]{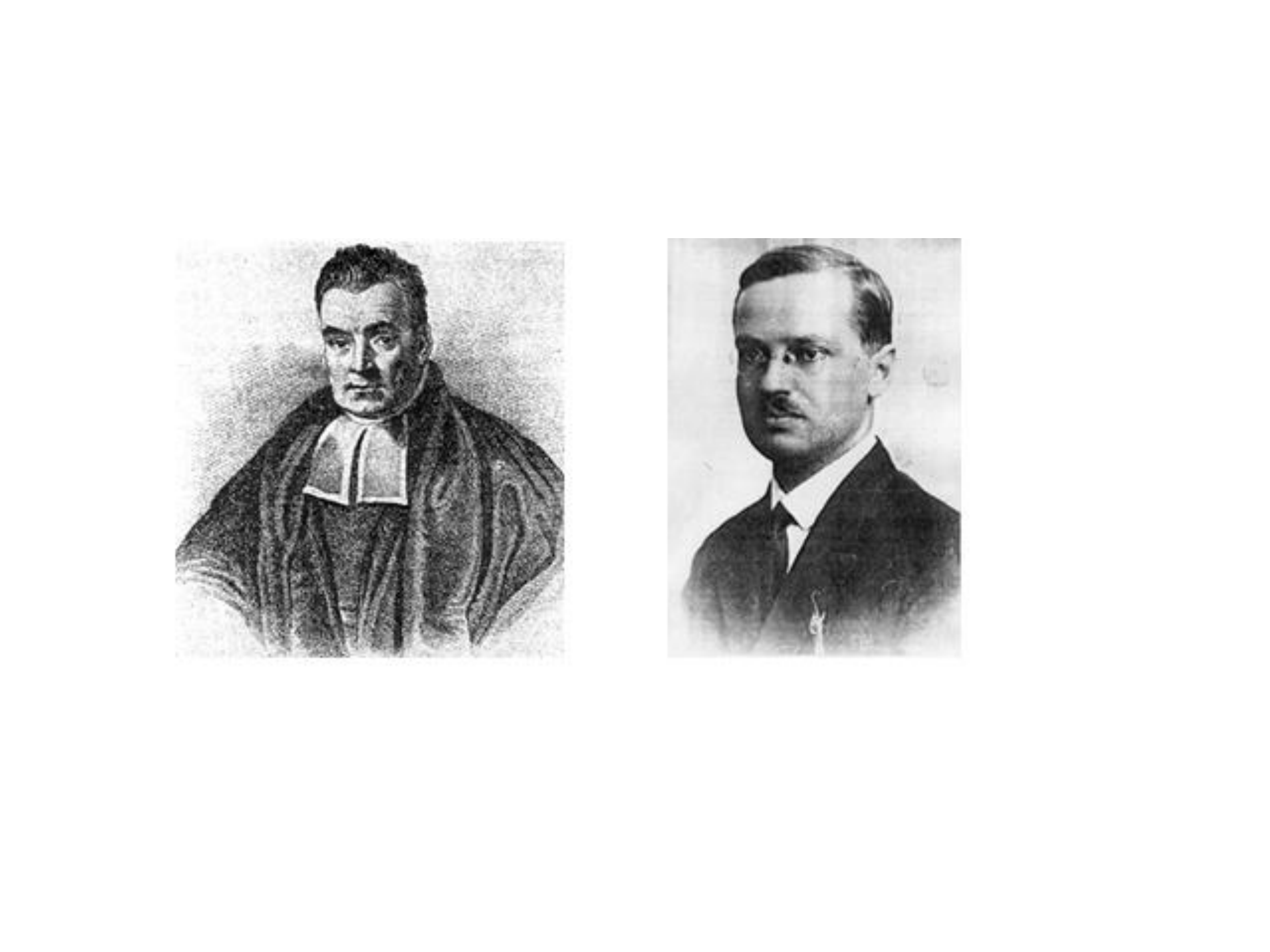}
\caption{ The Reverend Bayes (left), whose paper on his theorem was published posthumously 
in 1763; and Jerzy Neyman, a Polish statistician who played a crucial role in the development of the frequentist 
approach.
\label{fig:BayesNeyman}   }
\end{center}
\end{figure}


In this article, the fundamental differences between these two approaches
will be explained, and illustrated with examples from Physics and from
every-day life. We consider them in situations where we are trying 
to measure a parameter (e.g. the mass of the top quark), or are testing 
hypotheses (e.g. do we have evidence for the existence of the Higgs boson?)

\subsection{Why the fuss?}
Given that there are these fundamentally different ways of analysing data, 
how is it possible that many scientists spend a lifetime measuring all 
sorts of physical parameters, without being aware of the sharp differences 
of philosophy between the Bayesian and Frequentist approaches? The answer 
is that in the simplest of problems the two methods (and others too, like 
$\chi^2$ or maximum likelihood) can give the identical answer, that the 
probability that a parameter $\mu$ lies in the range $\mu_l$ to $\mu_u$ is, say, 68\%. 
By the `simplest of problems', we mean that the measured value $m$ is 
Gaussian distributed about the true value $\mu$ with known variance $\sigma^2$, 
and that $\mu$ can in principle have any value from minus infinity to plus infinity. 

However, in many practical problems in Particle Physics, these conditions are not
satisfied. The parameter may be restricted in range (masses cannot be negative),
and the data distribution may not be Gaussian (counting experiments often follow
the Poisson distribution). So there is ample opportunity for the results of 
Bayesian and frequentist analyses to differ. 
The two types of statisticians have often had strong 
criticisms of each other's approach.

\subsection{Probability}

The differences between the Bayesians and Frequentists start with their
interpretation of `probability'. Underpinning both of these is the mathematical 
approach, which is largely due to Russian mathematicians such as Kolmagorov. 
It is based on axioms (e.g. probability is a number in the range 0 to 1; the 
sum of the probabilities for something to occur and for it not to occur is
1; etc.). This is very important for manipulating probabilities, but provides 
little physical intuition about the concept. 

For frequentists, the probability $p$ of `something'  is defined in terms of 
a large number $N$ of essentially identical, independent trials: if the specified 
`something' happens in $s$ of these trials, $p$ is defined as the limit of the 
ratio $s/N$, as $N$ tends to infinity. Thus the probability of the sum of the 
numbers on two rolled dice adding up to 10 can be determined in this way
to be 1/12.

Bayesians attack this definition, as it  requires a large number of 
`essentially identical' trials. They claim that to determine whether the 
trials are indeed `essentially indentical' requires the concept of 
probability, and hence the definition is circular. 

Given that a repeated series of trials is required, frequentists are unable 
to assign probabilities to single events. Thus, with regard to whether it was 
raining in Manchester yesterday, there is no way of creating a large number of
`yesterdays' in order to determine the probability.  Frequentists would say
that, even though they might not know, in actual fact it either was raining or 
it wasn't, and so this is not a matter for assigning a probability. And the same 
remains true even if we replace `Manchester' by `the Sahara Desert'.

Another example would be the unwillingness of a frequentist to assign a 
probability to the statement that `the first astronaut to set foot on Mars 
will return to Earth alive.' This does not mean it is an uninteresting 
question, especially if you have been chosen to be on the first manned-mission
to Mars, but then, don't ask a frequentist to assess the probability.

 
A different type of example involves physical constants. Frequentists will
also not assign probabilities to statements involving the numerical values 
of physical parameters e.g Does dark matter constitute more than 25\% of the 
the critical density for our Universe? This again is a situation which cannot be checked 
by replicated tests. And again, it is either true or false, and not 
suitable for frequentist probabilities. A similar argument applies to 
statements about theories: a Frequentist will not allow probability 
assignments as to whether the Higgs boson exists.  

Bayesians have a very different approach. For them, probability is a personal
assessment of how likely they think something is to be true. It depends 
on their own judgement and/or previous knowledge about the situation, and can 
hence vary from person to person. Thus if I toss a coin, and ask you what is 
the probability of the result being heads, you are likely to say 50\%. But 
maybe I cheated and looked at the coin, and saw that it was tails, so for 
me the probability of heads is 0\%. Or maybe I just gave it a quick glance, 
and think (but am not certain) that it was tails, so I assign a probability 
of 20\% to heads.

Because Bayesians have this personal view of probability, they would be 
prepared to give numerical estimates for `one-off' situations (e.g. who 
gets this year's Nobel Prize?), for parameter values (e.g. fraction of dark 
matter), or concerning theories (e.g. existence of Higgs boson). Again, these 
numerical assessments could vary from person to person.

\vspace{0.2in}

\fbox{
\parbox {0.9\linewidth}  {
PERSONAL PROBABILITIES

This is a story I originally heard from Nobel Prize winner Frank Wilczek
in a slightly different context, but it illustrates the way that for Bayesians
the assessment of probability can differ from person to person.

A shy postdoc is attending a workshop on the topic of `Extra Dimensions'. 
Each evening, after an intensive day's work, he goes to the local bar, sits next 
to an empty chair and orders two glasses of wine, one for himself and the other
for the empty chair. By the third evening, the barman's curiosity cannot be 
controlled and he asks the postdoc why he always orders the extra glass of wine.
`I work on the theory of extra dimensions', explains the postdoc, `and it is
possible that there are beautiful girls out there in 12 dimensions, and maybe 
by quantum mechanical tunneling they might appear in our 3-dimensional world,
and perhaps one of them might materialise on this empty chair, and I would 
be the first person talking to her, and then she might go out with me'. `Yes',
says the barman, `but there are three very attractive real girls sitting over there 
on the other side of this bar. Why don't you go and ask them if they would 
go out with you?' `There's no point', replies the postdoc, `that would be 
very unlikely.' }} 
 
\vspace{0.2in}

It sounds as if this is very personal and not conducive to numerical estimates.
But Bayesians' assessment of probability should be consistent with the 
`fair bet' concept. If a Bayesian believes that a certain statement has a
10\% probability of being true, they should be prepared to offer odds of 9 to 1
(or 1 to 9) to someone who is prepared to bet with them on this being true (or
false, respectively). With a poor assessment of the probability, they would be 
in danger of losing money.
 
\section{LIKELIHOODS, BAYES THEOREM AND PRIORS}
We now have a relevant digression into considering likelihood functions,
and then introduce Bayes Theorem and Priors, essential ingredients of the 
Bayesian approach.

\subsection{Likelihoods}
\label{likelihoods}
The likelihood approach is a very powerful one for parameter determination, and 
is also very much involved in Bayesian and Frequentist methods for this. 
Likelihood ratios are also used for checking which of two theories provides a 
better description of the data.

The likelihood function is best illustrated by a simple example. Imagine we 
are performing a counting experiment for some fairly rare process. For example.
we may be interested in the flux $\mu$ of cosmic ray showers with energies above $10^{20}$ 
electron volts. We have a large detector of known area, and find $n_0$
high energy showers 
(e.g. 2) when running the detector for one year. We want to make a statement 
about the value of the actual flux $\mu$  and its uncertainty.

Assuming these cosmic rays are falling on earth at a constant rate, and 
are independent of each other, if the true rate is $\mu$, the conditional 
probability $P(n|\mu)$ of obtaining n counts is given by the Poisson 
distribution as
\begin{equation} 
           P(n|\mu) = e^{-\mu}  \mu^n/n!  
\label{eqn:pdf}
\end{equation}
Then the likelihood is defined by replacing n in the above formula by
the observed value $n_0$. i.e.
\begin{equation}
           L(\mu|n_0) = e^{-\mu}  \mu^{n_0} /n_0! 
\label{eqn:likelihood}
\end{equation}
This likelihood is regarded as a function of $\mu$, for the fixed data value
$n_0$. (For example, if we observe 2 events, the likelihood is $\mu^2 \, e^{-\mu}/2$.)  
It is the probability of observing the data, for different choices of $\mu$.
Then the likelihood estimate of a parameter $\mu$ is that which maximises 
the likelihood i.e. It is the value of $\mu$ which maximises the probability 
of observing the actual data $n_0$. (In our case, not surprisingly the 
likelihood estimate of $\mu$ is simply $n_0$.) Values of $\mu$ for which the likelihood 
is small are regarded as excluded, and the uncertainty on $\mu$ is related to the 
width of the likelihood distribution.

\vspace{0.2in}

\fbox{
\parbox {0.9\linewidth}  {
A POISSON PUZZLE?

According to the Poisson distribution, if the expected number of observations in a 
specified time is $\mu$, the probabilities $P(1|\mu)$ and $P(2|\mu)$ are
\begin{equation*}
P(1|\mu) = \mu\, e^{-\mu}, \ \ \   P(2|\mu) = \mu^2\, e^{-\mu}/2  
\end{equation*}         
For small $\mu$, these are approximately $\mu$ and $\mu^2/2$ respectively. Given 
the fact that the probability for observing one rare event in the time interval is
$\mu$, why is the probability for observing two independent events equal to $\mu^2/2$, 
rather than simply $\mu^2$, as perhaps expected from eqn. (\ref{independence})?
}}

\vspace{0.2in}

\begin{figure}[h]
\begin{center}
\includegraphics[width=.7\textwidth]{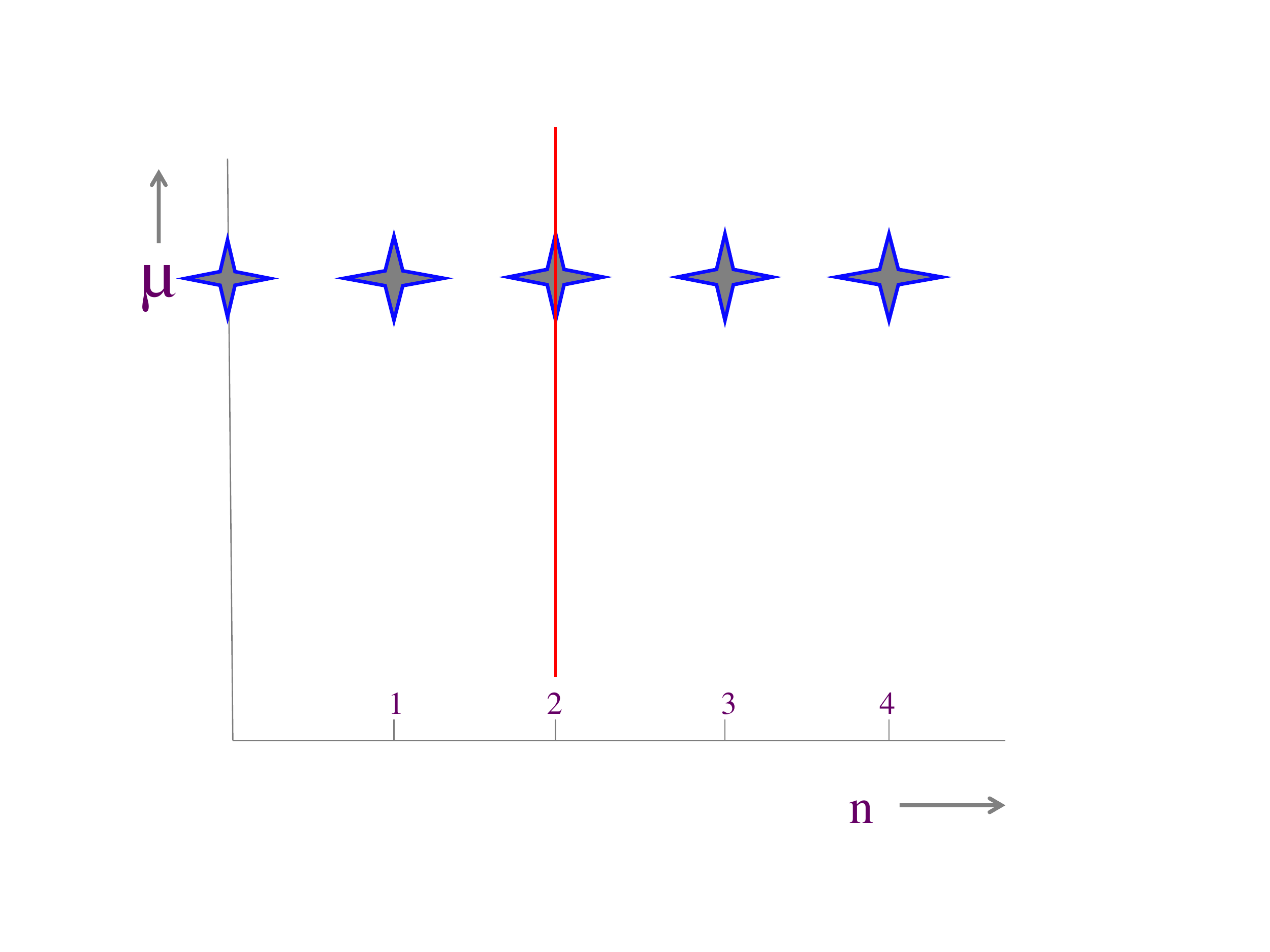}
\caption{Illustration of the difference between the probability density distribution for the
integer variable $n$ and the likelihood function for the continuous parameter $\mu$, for the 
Poisson distribution (see eqns. (\ref{eqn:pdf}) and (\ref{eqn:likelihood})). They involve 
the same function of $n$ and $\mu$, but it is evaluated
at fixed $\mu$ for the $pdf$, but at fixed $n$ for the likelihood.
\label{fig:pdf_andL}    }
\end{center}
\end{figure}

It is really important not to confuse the Poisson probability $P(n|\mu)$ with the 
likelihood function $L(\mu|n_0)$, even though eqns. (\ref{eqn:pdf}) 
and (\ref{eqn:likelihood}) bear a 
remarkable similarity\footnote{The `!' symbol in eqns (\ref{eqn:pdf}) and
(\ref{eqn:likelihood}) not only expresses surprise (`Wow! These equations look very 
similar), but it also denotes the factorial.} . 
The distinction should be easy in this case: $P(n|\mu)$ is a 
function of the discrete variable $n$ at fixed $\mu$, while $L(\mu|n_0$) is a function 
of the continuous variable $\mu$ at fixed $n_0$ (see fig. \ref{fig:pdf_andL}).
Furthermore, $P(n|\mu)$ are real 
probabilities,
while the likelihood $L(\mu|n_0)$ cannot be interpreted as a 
probablity density 
(it does not transform as expected for a probability density if the parameter is chosen, 
for example,  as $1/\mu$ rather than $\mu$).


\subsection{Bayes theorem}  
If we consider two `events' $A$ and $B$ (in the statistical sense), we can write the 
probability $P(A\ and\ B)$ of them both happening as
\begin{equation}
         P(A\ and\ B) = P(A|B)\,  P(B)
\label{PAandB}
\end{equation}
where $P(A|B)$ is the conditional probability of $A$ happening, given the fact that
$B$ has occurred. An example could be where we select a random day from last year,
and $A$ is whether it was snowy in Oslo, and $B$ that it was a December day. Then Bayes 
Theorem says that the probability of choosing a snowy December day is equal
to the probability of it being snowy in December, multiplied by 31/365 (the probability 
of a random day being in December). If the probability of $A$ occuring does not depend on
whether $B$ has done so, eqn. (\ref{PAandB}) reduces to the better-known result that 
\begin{equation}
              P(A\ and\ B) = P(A) \, P(B), \ \ \  for\ independent\ A\ and\ B
\label{independence}
\end{equation}

\vspace{0.2in}

\fbox{
\parbox {0.9\linewidth}  {
 CONDITIONAL PROBABILITY

Conditional Probability $P(A|B)$  is 
the probability of $A$, given the fact that $B$ has happened. For example, the probability of obtaining a 
4 on the throw of a dice is 1/6; but if we accept only even results, the conditional probability for a 4,
given that the number is even,  is 1/3.
}}

\vspace{0.2in}

Because $P(A\, and\,  B)$ is symmetric in $A$ and $B$,
\begin{equation}
         P(A\ and\ B) = P(A|B)\,  P(B) = P(B|A)\,  P(A)
\end{equation}
Then Bayes Theorem is derived from the second equality above:
\begin{equation}
         P(A|B) = P(B|A)\,  P(B)/P(A)
\end{equation}
i.e. It relates $P(A|B)$ to $P(B|A)$. (See section \ref{PABnePBA} for examples where these are 
obviously not equal.)

It should be stressed that Bayes Theorem itself is not controversial, and
frequentists are willing to make use of it, provided the various probabilities
are genuine frequentist ones. The controversy begins when Bayesians replace $A$ by 
a theoretical parameter (and $B$ is the observed data). The theorem then
states that 
\begin{equation}
       P(param|data) \propto  P(data|param) \times  P(param)
\end{equation}
where $P(data|param)$ is just the likelihood function; $P(param)$  is the Bayesian prior
density, and expresses what was known about the parameter before our measurement;
and $P(param|data)$ is the Bayesian posterior probability density for the parameter,
and contains the information about the parameter obtained by combining the
prior information with that from our measurement.

\vspace{0.2in}

\fbox{
\parbox {0.9\linewidth}  {
BAYESIAN POSTERIORS

Jim Berger says that he and his wife have professions
that are similar, but with a small difference. He is a Bayesian Statistician
and she is a fitness trainer. The similarity is that they both aim to 
optimise posteriors, but while he wants to maximise them, she aims to 
minimise her clients'  posteriors.
}
}

\vspace{0.2in}

The frequentist objection to this is that the prior and the posterior both refer
to parameter values; while this is allowed for Bayesians, it is strictly forbidden
in the frequentist approach. In addition to this, complications are caused by the 
need to choose a probability density for the prior.

\subsection{Bayesian priors}
\label{priors}
In order to obtain the Bayesian posterior probability distribution from the
likelihood function, the latter needs to be multiplied by the Bayesian prior.
There are several flavours of Bayesians, who have different motivations for 
their choice of prior. In this article, we will concentrate on 
evidence-based priors. So if in our Poisson example of Section \ref{likelihoods}, there was a previous 
measurement of $\mu$ which gave the result $5\pm1$, the prior might be chosen as
a Gaussian in $\mu$, centred on 5 with standard deviation 1 (and probably
truncated at zero). Then the posterior, assuming 2 observed counts, would be
\begin{equation}
         P(\mu|n_0 = 2) = (\mu^2\, e^{-\mu}/2) \times ( e^{-(\mu-5)^2/2}/\sqrt{2\pi}),
\end{equation}
where the first factor on the right  is the likelihood $L(\mu|n_0 = 2)$, and the 
second is the prior $\pi(\mu)$.

This is all very well when previous data on $\mu$ exists. But what if our 
measurement is ground-breaking, and essentially nothing is previously
known about $\mu$? How do we now choose the prior $\pi(\mu)$? The `obvious'
answer is to choose a prior that is independent of $\mu$ (but zero for 
unphysical negative $\mu$), so as not to favour any particular value of
$\mu$. However, do we really believe, as implied by the constant prior, 
that $\mu$ is as likely to be in the range $10^{500}$ to $10^{500} + 0.5$, as in 0.1 to
0.6?

Another problem is that if we are aiming to use a flat prior to express
our ignorance about a parameter, it is not clear why we should 
choose one functional form for the parameter rather than another. For example,
if we are aiming to provide a very precise measurement of the mass $m$ of the
tau neutrino, should we parametrise our ignorance of its mass by a flat
prior in $m$, $m^2$, $\ln(m)$, etc? Basically priors may be not bad for 
parametrising prior knowledge, but are not so good for prior ignorance.

\subsection{$P(A|B) \ne P(B|A)$}
\label{PABnePBA}
Bayes Theorem relates the conditional probabilities  $P(A|B)$ and $P(B|A)$. 
People often confuse these two
probabilities, and may erroneously think they are the same. Thus
journalists or even Laboratory Spokespersons may incorrectly say that there 
is a 99.9\% probability that some particle exists, rather than the correct statement that
under the null hypothesis that it does not, the data are very unlikely.

A convincing example of their difference is provided by a database containing 
just 2 pieces of information about everyone on Earth: their sex and whether or not 
they are pregnant. We extract a random person from the database, who turns out 
to be female. Given that the person is female, the chance of being pregnant is about
3\%. We then extract another random person, who turns out to be pregnant. Given the 
fact that the person is pregnant, the probability that they are female is 100\%. i.e.
\begin{equation}
           P(pregnant|female) \ll P(female|pregnant)     
\end{equation}

Similarly, if you select a card randomly from a deck of 52, the probability of it being  
an ace, if it  happens to be a spade, is 1/13; however, the probability 
of a spade, given that it is an ace, is 1/4.

\subsection{A Bayesian example}
Imagine that you, a Bayesian, are betting on the results of coin flips. Each time you 
bet `Heads', and for the first 5 flips it comes down `Tails'. Given that the
probability of being wrong 5 times is 3\%, should you suspect it is not a fair coin?

We regard this as a parameter-estimation problem, and want
to see whether the probability $p_H$ of `Heads' is consistent with 0.5. The data (no
heads in 5 spins) enables us to calculate the likelihood function, but in
order to extract the posterior probability as a function of $p_H$, we must multiply
the likelihood by a prior $\pi(p_H)$. Now if the person betting against us is a
complete stranger, we might assign a constant value for $p_H$ in the range 0 to 1; then the 
posterior is such that $p_H\ =\ 0.5$ looks unlikely. On the other hand, if it is our 
local village priest, we are so convinced that he is honest, we use a delta 
function at $p_H\ =\ 0.5$, and then even if the coin continues to fall down `Tails', we will
still believe that it is fair. Thus our conclusion depends very much on which prior we 
choose.

Given the freedom to select one's prior, it seems as if Bayesian intervals 
for a parameter can be very dependent on this choice.
But in some cases, the `data overwhelms the prior',
and the result becomes very insensitive to the choice of prior. For example, the mass
of the intermediate vector boson ($Z^0$) was measured at the LEP (Large Electron Positron)
Collider at CERN. The result was that the likelihood function was essentially a 
Gaussian at 91,188 $MeV/c^2$, with a width of 2 $MeV/c^2$. A Bayesian now has to multiply
this by the prior probability density for the $Z^0$ mass. However, any reasonable 
choice of prior will vary very little over the range of a few parts in $10^5$, 
and so in this case the posterior is essentially independent of the prior.

\section{PARAMETER DETERMINATION:  BAYESIAN APPROACH}
We illustrate the Baysian approach using a simple example of the determination of 
the lifetime of some radioactive material. The probability density $p$ for a decay at 
time $t$ is given by
\begin{equation}
             p(t|\tau) = (1/\tau)\,  e^{-t/\tau}
\label{pttau}
\end{equation}
where $\tau$ is the lifetime we want to estimate. We can estimate $\tau$ from a set 
of decays at observed times $t_i$. To simplify the problem we assume 
we have only one decay at time $t_1$ (which will not give us a very accurate estimate 
of $\tau$).

The likelihood is
\begin{equation}
L(\tau) \ = \ (1/\tau)\,  e^{-t_1/\tau},
\end{equation}
and we have to multiply this by our choice of prior for $\tau$, to obtain the posterior 
$p(\tau|t_1) \ = \  L(\tau)\,  \pi(\tau)$. As usual, there is a choice for $\pi(\tau)$ of
an evidence-based prior derived from a previous measurement (in which case our
posterior and the resulting range for $\tau$ will be based not only on our 
measurement, but also on the previous one), ignorance, theoretical motivation, etc.
Because in many cases the choice of prior is not unique, Bayesian analyses 
are supposed to present results for several plausible priors, so as
to investigate the sensitivity of the result to the choice of prior. 

Once the posterior is available, several options are available for determining a range of preferred
$\tau$ values at some chosen probability level $\gamma$, i.e. 
\begin{equation}
\int_{\tau_l}^{\tau_u} \! p(\tau|t_1) \, \mathrm{d} \tau  = \gamma .
\label{interval}
\end{equation} 
Possibilities include:
\begin{itemize}
\item{A central range from $\tau_l$ to $ \tau_u$ could be obtained by having probabilities of
 $(1 - \gamma)/2$  below the range,  and $(1 - \gamma)/2$ above it.}
\item{The  upper limit $\tau_{UL}$ is obtained by setting the limits of integration in eqn. (\ref{interval})
from zero to $\tau_{UL}$.}
\item{In a similar manner, a lower limit $\tau_{LL}$ is obtained, using integration limits  $\tau_{LL}$
and infinity.}
\item{The shortest posterior range in $\tau$ containing probabability $\gamma$ is also popular, but 
does not correspond to the shortest range in the decay rate $1/\tau$, or for other 
reparametrisations of the variable of interest.}
\end{itemize}

\section{PARAMETER DETERMINATION:  FREQUENTIST APPROACH}
We now consider the frequentist approach for the same problem as in the previous section.


\begin{figure}[h]
\begin{center}
\includegraphics[width=.9\textwidth]{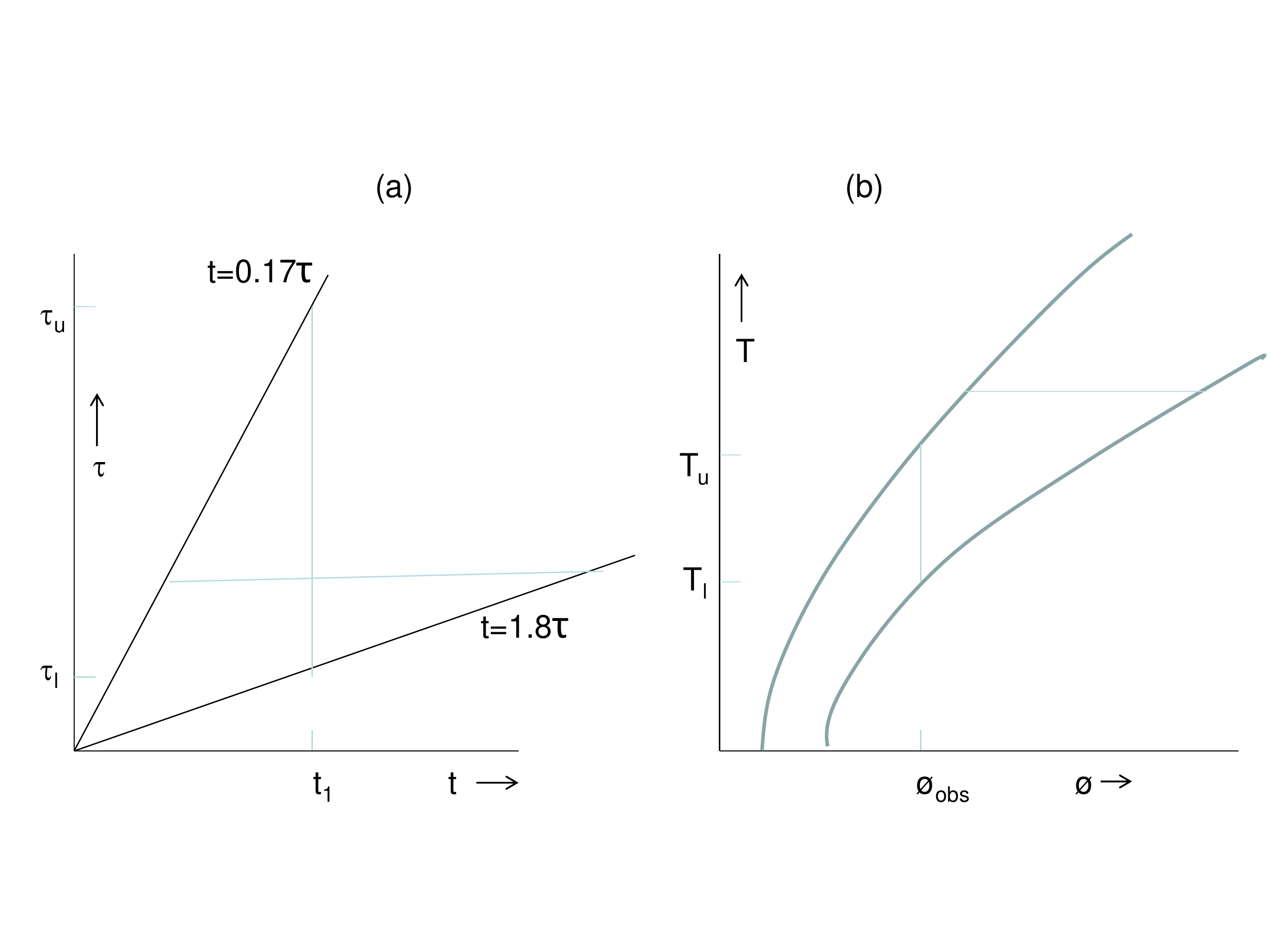}
\caption{
The Neyman construction. 
(a) For the exponential parameter $\tau$,
the central confidence  band  between the lines $t \ = \  0.17\tau$ and $t \ = \ 1.8\tau$  gives the likely 
values (at the 68\% 
level) of $t$ for each $\tau$. Then a vertical line at the observed $ t_1$ intersects 
the edges of the confidence  band at $\tau_l$ and $\tau_u$, and these define the 
frequentist range for $\tau$. 
(b) Here the theory parameter is the temperature $T$ at the centre of the sun,
and the data $\phi$ is the measured flux of solar neutrinos, both in arbitrary units.
A measurement of the flux $\phi_{obs}$  determines a range of temperatures ($T_l$ to $T_u$) at the sun's centre.
\label{fig:Neyman} }
\end{center}
\end{figure}

The Neyman construction is used to show on a plot of the parameter $\tau$ 
versus the data $t$ the likely values of $t$ for each $\tau$ (see 
fig. \ref{fig:Neyman}(a)).
This is achieved by using $p(t|\tau)$ of eqn. (\ref{pttau}) for a given $\tau$ to select a region 
of $t$ such that the integral of $p(t|\tau)$ over this range of $t$ is, say, 68\%\footnote{This 
definition does not provide a unique range. The one we show 
has a probability of 16\% on either side of the shaded region, which is then known
as a central interval. An alternative would be to have the whole of the 32\% 
on the left hand side of the confidence interval; this would be useful for 
producing upper limits on $\tau$.};
an example is denoted by the horizontal line in the figure. By 
repeating this procedure for all $\tau$, we obtain the `confidence band'. In our example, 
the edges of the band are defined by the straight lines
$t = 0.17\tau$ and $t = 1.8\tau$.  Finally 
we use the actual observed value $t_1$ to read off the range of $\tau$ values 
($\tau_l$ to $\tau_u$, which are $t_1/1.8$ and $t_1/0.17$ repectively) for which $t_1$ is 
a likely observation. For larger values of 
$\tau$, $t_1$ is too small to be likely, and similarly for smaller $\tau$, $t_1$ is 
too large. 

In a more plausible scenario where the data consisted of a set of observed 
decay times $t_i$, the data statistic could be the mean of the $t_i$. Then the 
confidence band would be narrower than in figure, and the range of acceptable 
$\tau$ values would be shorter. 

An important feature of this construction is that it does
not require a prior distribution for $\tau$, thus avoiding the possible ambiguity that
that would have introduced. Another point to note is that the
frequentist approach does not claim that the range $\tau_l$ to $\tau_u$ is
probable. Nor does it make any statement about different values within 
this range; it is merely that this is the range of $\tau$ values for which 
the observed data is likely (at the chosen confidence level).

Fig. \ref{fig:Neyman}(b) shows a more interesting example. An experiment aims to measure the
temperature $T$ of the fusion reactor at the centre of the sun, by using a month's
running of a solar neutrino detector to estimate the neutrino flux $\phi$ from the sun. 
Assuming we know all about fusion processes and convection in the sun, the properties
of neutrinos, the performance of our detectors, etc, we can construct at each $T$ 
a region in $\phi$ such that there is a 68\% probability the experimental result 
would lie in it. Then we use the actual measured flux $\phi_{obs}$ to determine 
the accepted range for $T$.

\subsection{Coverage}
\label{Coverage}
For repetitions of an experiment using a particular statistical analysis to determine a range for 
the parameter of interest, where the data sets differ from each other just by 
statistical fluctuations, the coverage is the fraction of the parameter's  intervals that contain 
the true value of the parameter. This can be determined from Monte Carlo simulation, 
or in some simple cases analytically. 
Coverage is a property of the statistical technique that is used to construct the 
intervals, and does not apply to a single measurement. 

Techniques for which the coverage is equal to the nominal value (e.g. 68\%) 
for all values of the parameter are said to have exact coverage.  If the coverage 
drops below the nominal value, the method is said to under-cover. Frequentist regard this as
bad: if the actual coverage for determining the parameter is only 30\% rather than 
the nominal 68\%, just quoting the range for the parameter as determined by 
that method is likely to mislead a reader into believing that your result is more 
accurate than it really is. Over-coverage does not have this problem, but it suggests
that maybe the confidence intervals produced by that method are more 
conservative (i.e. wider) than they need be.

A particularly important property of the Neyman construction is that the 
confidence intervals for the parameter have the property of not
undercovering. This property is not guaranteed for other techniques (e.g. Bayesian,
$\chi^2$, maximum likelihood, method of moments).


\begin{figure}[h]
\begin{center}
\includegraphics[width=.9\textwidth]{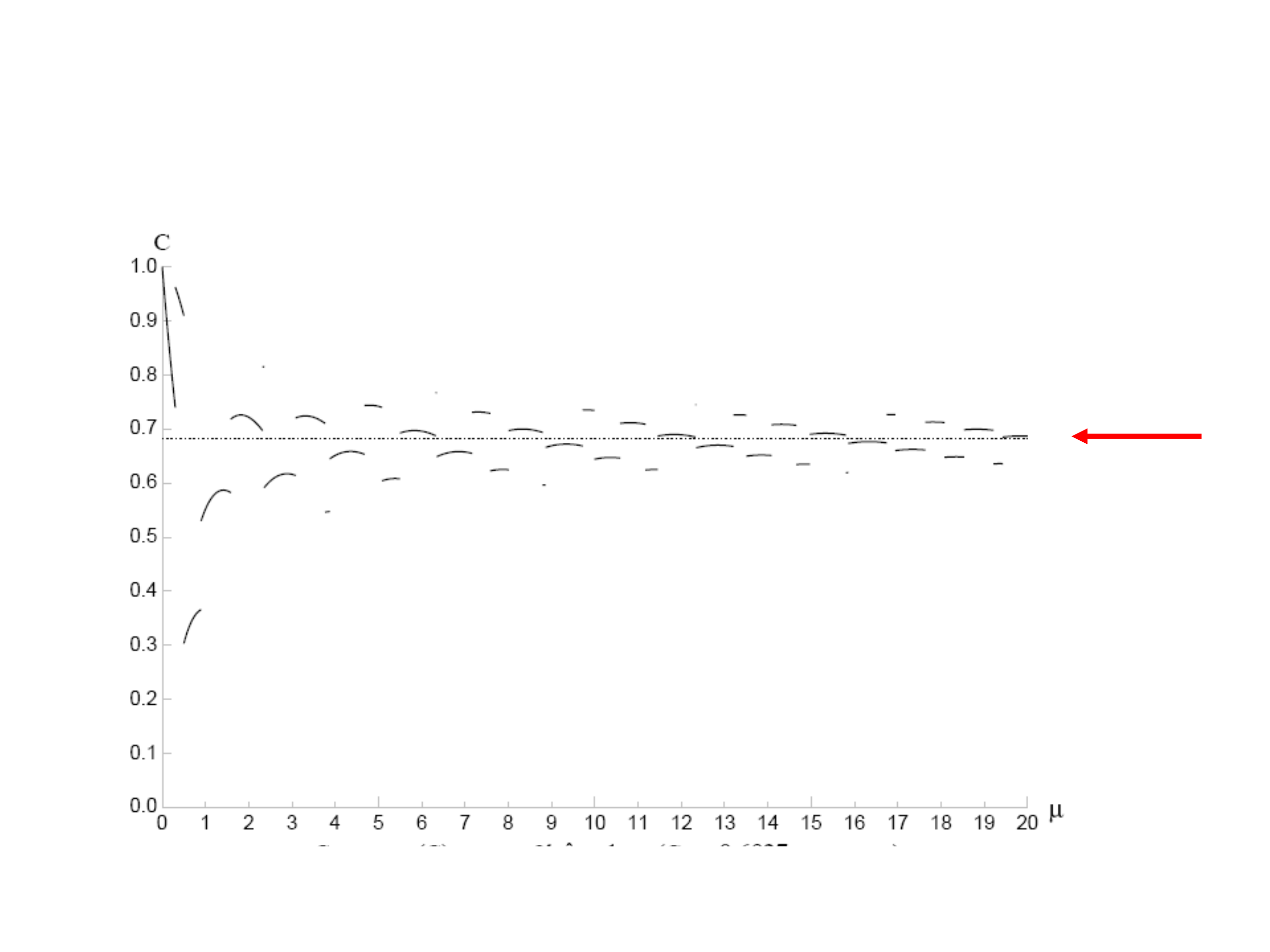}
\caption{Coverage $C$ for Poisson parameter intervals, as determined by the $\Delta(\ln(L)) =0.5$ rule. 
Repeated trials (all using the same Poisson parameter $\mu$) yield different values of the observation $n$, each 
resulting in a range $\mu_l$ to $\mu_u$ for $\mu$; then  $C$ is the fraction of trials that give ranges which 
include the value of $\mu$ chosen for the trials. The coverage $C$ varies with $\mu$, and has 
discontinuities because the data $n$ can take on only discrete integer values. For large $\mu$, $C$ 
seems to approach the expected 0.68.   
\label{fig:PoissonCoverage}}
\end{center}
\end{figure}

Fig. \ref{fig:PoissonCoverage} shows the coverage $C$ for the following situation. An 
experiment is performed to determine the rate $\mu$ of some Poisson counting experiment,
and $n$ counts are observed. The statistical procedure chosen for determining a  68\% 
range for $\mu$ is the likelihood method with the $\Delta(lnL) =0.5$ rule to define
the ends of the range. In envisaged repetitions of the experiment, $n$ will vary 
according to a Poisson distribution with mean $\mu_0$. Then $C(\mu_0)$ is the fraction of the
resulting ranges for $\mu$ which include $\mu_0$. The likelihood method does not have the frequentist
guarantee of coverage, and indeed large under- and over-coverage occur, especially at
low $\mu$\cite{Coverage}. 

\section{PARAMETER DETERMINATION: COMMON ISSUES}
Here we discuss some issues that are common to both Frequentist and Bayesian approaches. 

\subsection{Parameters with limited range}
Very often a physical parameter has meaning only over a limited range. For example,
the square of the mass of the neutrino  ($m_\nu^2$) produced in nuclear beta decay cannot be 
negative, the branching ratio for some particular decay mode of an elementary particle
must be between zero and one, etc. Bayesians can incorporate this information by 
setting the prior for the parameter to zero in the non-physical region. This ensures that
the best estimate of a parameter or an upper limit for it are guaranteed to be physical.
In contrast, a frequentist upper limit could well turn out to be unphysical, or 
the range for $m^2_\nu$ could be empty (i.e. there was no physical value of $m_\nu^2$ for 
which the data was likely); in general Particle Physicists are unhappy with such a situation. 

For many years, experiments estimating $m^2_\nu$ had `likelihood functions' that 
maximised at negative values. Upper limits for $m^2_\nu$ were then usually derived 
by Bayesian methods.

\subsection{Interpretation of  $\mu_u \ge \mu\ge \mu_l$}
Both Bayesian and Frequentist methods of parameter determination end up with a
statement of the form  $\mu_u \ge \mu \ge \mu_l$ at some probability level, but their 
interpretations are very different. 

For frequentists, the parameter $\mu$ is unknown, 
but it does have a true value and, as discussed earlier, it is not suitable for 
probability statements. So the probability refers to the range $\mu_l$ to $\mu_u$. If the
experiment were to be repeated many times, a series of ranges for $\mu$ would be obtained,
and the probability refers to what fraction of these ranges contain the true value; 
this is just the coverage mentioned in Section \ref{Coverage}. Thus frequentists regard the ends of 
the range as random variables.

For Bayesians, $\mu_u$ and $\mu_l$ have been detemined by the experimental analysis, and 
are considered fixed; Bayesians do not want to be involved in deciding what would 
have happened in hypothetical repetitions of the experiment. But they are prepared to treat 
the unknown physical constant as if it were a random variable, and for them the 
probability refers to the fraction of the Bayesian posterior probability density for $\mu$ is within
the quoted range.

\begin{table} 
\caption{Interpretations of ``$\mu_u \ge  \mu \ge \mu_l$ at 68\% confidence level"}
\begin{center}
\begin{tabular} {|p{3.5cm}|p{5cm}|p{5cm}|}
\hline
                &    Bayes   &                    Frequentist \\
\cline{2-3}
What is fixed?   &  $\mu_u, \mu_l$    &                  $\mu$  \\
What is variable?  &    $\mu$             &           $\mu_u, \mu_l$  \\
What does 68\% probability apply to?    &  Single measurement:  percentage of $\mu$'s posterior in range &     Set of measurements:   percentage of ranges  $\mu_l \rightarrow \mu_u$  that contain $\mu$ \\
\hline
\end{tabular}
\end{center}
\end{table}

\subsection{Dealing with systematics}
Very often, in trying to estimate a parameter, some other quantity involved in the 
analysis is not known exactly, and this can affect the deduced range for the parameter 
of interest. For example,
in the original Reines and Cowan experiment \cite{Reines} to discover the electron neutrino,
a detector sensitive to neutrinos interacting in it was built close to a powerful 
nuclear reactor. However, there were also background processes which mimic the 
interactions of the reactor neutrinos. Then the observed number of counts $n$ is likely 
to be Poisson distributed with mean $b + s$:
\begin{equation}
                   P(n) = e^{-(b+s)}\, (b+s)^n /n!
\end{equation}
where $b$ is the expected background, and $s$ is the signal rate.
If $b$ is precisely known, $s$ is the only unknown parameter, and can be determined essentially
as described earlier. But if there is some uncertainty in the expected value of $b$, this
results in a systematic uncertainty in the answer. Statisticians refer to $b$ as a 
nuisance parameter.                        

Bayesians tend to treat all parameters (i.e. those of physical interest and nuisance 
parameters) in a similar manner. Thus, assuming that the background $b$ has been estimated
in a subsidiary counting experiment as $m_0$ while the result of the main measurement of 
$s + b$ was $n_0$, they would start by writing the likelihood for $s$ and $b$ as
\begin{equation}
           L(s,b|n_0,m_0) = (e^{-(s+b)}\, (s+b)^{n_0}/n_0!) \times (e^{-b}\, b^{m_0}/m_0!)
\end{equation}
Next this is multiplied by the chosen prior $\pi(s,b)$ for $s$ and $b$, to give
the posterior probability $p(s,b)$ for the parameter of interest $s$ and the nuisance 
parameter for the background $b$. Then this is integrated (or `marginalised') over $b$
to give the probability density just for the parameter of interest:
\begin{equation}
          p(s) = \int \!  p(s,b)\ \mathrm{d} b
\end{equation}
Finally the required parameter range is extracted from $p(s)$ e.g. a central 68\% range.

In contrast, frequentists start from the probability density $p(n,m|s,b)$
for observing any $n$ and $m$ as 
\begin{equation}
         p(n,m|s,b) = (e^{-(s+b)}\,  (s+b)^n/n!)) \times (e^{-b}\, b^m/m!)
\end{equation}
The fully frequentist method consists in performing a Neyman construction to produce a 
confidence belt for likely data $(n,m)$ as a function of the parameters $(s,b)$. In
analogy with the simpler problems discussed earlier, the actual data ($n_0,m_0$)
is then used to read off the region in parameter space $(s,b)$ for which the data is likely. If 
a range just for $s$ is desired, it could be taken as the extrema of the $(s,b)$ region, 
although this will give rise to overcoverage. 

There are also various approximate methods, which are simpler than the full 
Neyman construction and which tend to produce less overcoverage (but for which
the frequentist guarantee of coverage no longer applies). An example is the
profile likelihood approach, in which the  probability $ p(n,m|s,b)$ is
replaced by $p_{prof}(n,m|s,b_{best}(s))$, where $b_{best}(s)$ is the value of $b$ which maximises
the probability for that value of $s$; because $b_{best}(s)$ is a function of $s$,
the profiled probability depends on the single parameter $s$, which simplifies the problem.

\vspace{0.2in}

\fbox{
\parbox {0.9\linewidth}  {
PROFILE LIKELIHOOD

In many situations, the probability of observing a particular set of data $d$ 
depends not only on a parameter of physical interest $\phi$ (e.g. the mass of the Higgs boson),
but also on some other so-called nuisance parameters $\nu$ (e.g. a scale
factor for correcting jet energies as measured in the detector). Then
the likelihood $L(\phi,\nu|d)$ is a function of both sets of parameters $\phi$ and $\nu$.
In order to draw conclusions about $\phi$, it is ofter helpful to consider the 
{\bf profile} likelihood $L_{prof}(\phi,\nu_{max}(\phi)|d)$, where for each value of 
$\phi$, the nuisance parameters are chosen to maximise the full likelihood
$L(\phi,\nu|d)$, i.e. $\nu_{max}$ varies with $\phi$. However, $L_{prof}$ now is 
a function just of $\phi$ but not of the nuisance parameters $\nu$, thereby simplifying
the problem of making inferences about the parameter of interest $\phi$, 
at the cost of losing some of the properties of the likelihood function.

Rather than maximising $L$ with respect to $\nu$, Bayesian methods tend to 
{\bf marginalise}, i.e. integrate the likelihood with respect to $\nu$, usually 
after using priors for $\nu$, to convert $L$ into a posterior 
probability distribution for $\phi$.

For the case where $L$ is a multi-dimensional Gaussian distribution such as
\begin{equation*}
L \propto exp\{-(a\phi^2\ + \ 2b\phi\nu \ + c\nu^2)\}, \nonumber
\end{equation*}  
marginalisation over $\nu$ or profiling with respect to it  will give the 
same functional form for the modified likelihoods.
}}

\vspace{0.2in}

Ref. \cite{HeinrichLyons} contains a longer discussion of systematics, while 
Demortier\cite{Luc_systematics} deals with ways of incorporating systematics in
both parameter determination and hypothesis testing. 
    
\vspace{0.2in}

\section{HYPOTHESIS TESTING}
\begin{figure}[h]
\begin{center}
\includegraphics[width=.9\textwidth]{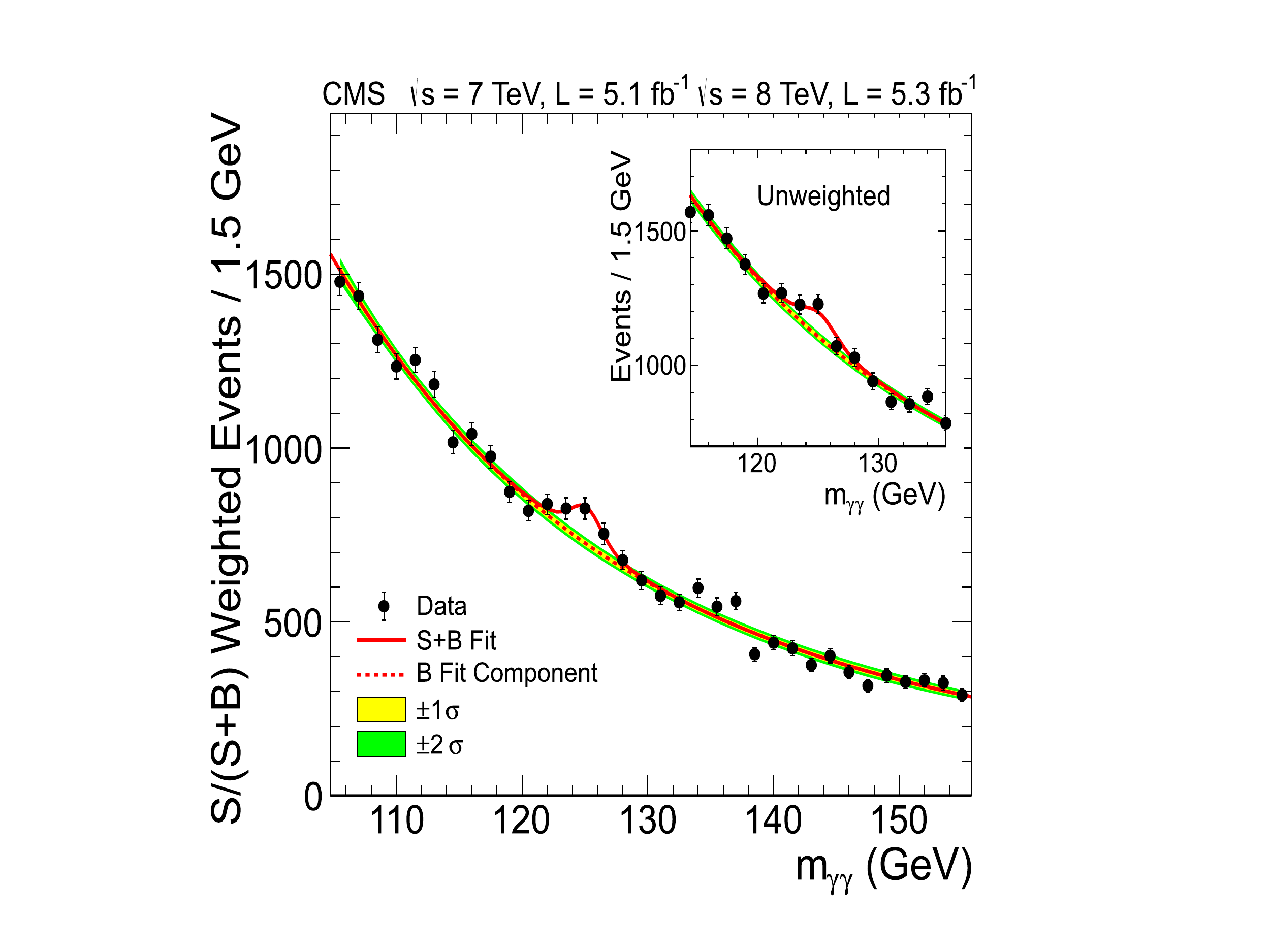}
\caption{The observed  distribution in the CMS experiment 
for the effective mass $m_{\gamma\gamma}$ of pairs of 
$\gamma$s produced in high energy proton-proton
collisions at CERN's LHC. If the Higgs boson exists and decays to a pair
of $\gamma$s, it could result in a peak centred on the mass of the Higgs.
Otherwise, the expected distribution is expected to be smooth. In the main plot, the events are 
weighted according to their quality; the inset shows unweighted events. The apparent 
peak around 125 GeV is part of the evidence for the existence of a new particle, whose 
properties seem consistent with those expected for a Higgs boson.
\label{fig:Higgs_mass_plot}}
\end{center}
\end{figure}

Possibly more interesting than Parameter Determination is Hypothesis Testing. Here the
issue is to decide which of two (or more) competing theories provides a better fit to
some data. For example, was data collected at the Large Hadron Collider at CERN in the 
first half of 2012 more consistent with what is known as the Standard Model (S.M.) of 
Particle Physics without anything new, or with the production of the Higgs boson 
in addition to the known S.M. processes? (See fig. \ref{fig:Higgs_mass_plot})


In Particle Physics, for reasons to be explained below, it is much more common to use a
frequentist method to decide. In other fields, Bayesian approaches tend to be favoured.
We discuss Bayesian methods briefly in section \ref{Bayesian_HT}.

\subsection{Frequentist approach}
\label{Fr_HT}

The first task is to choose some data statistic $t$ which will help distinguish
between the hypotheses. In the simple case of a counting experiment, where the data 
consists just of the number of accumulated counts $n_0$ for a given amount of running time,
it could simply be $n_0$. Then in most cases new physics would 
manifest itself in a larger number of counts when the expected rate is $s + b$, than 
if there were just background; here $s$ and $b$ are the expected signal and background 
rates respectively.

In more complicated cases, the data could consist of one or more histograms or 
multi-dimensional distributions. Then usually $t$ is chosen as a 
likelihood ratio for the data, assuming the two hypotheses:
\begin{equation}
              t = L_1(H_1|d)/L_0(H_0|d)
\end{equation}
where $L_1$ is the likelihood for $H_1$ (the hypothesis of signal + background), given 
the data, while $L_0$ is for the background only hypothesis $H_0$, given the same data.
When the hypotheses are completely specified without any free parameters, they are known
as `simple hypotheses' and the above formulation is satisfactory. Then the Neyman-Pearson 
lemma\cite{NP} says that if we choose $H_0$ based on the likelihood ratio being below some 
suitably defined cut-off, this will guarantee that we will achieve the lowest rate for ``Errors of the 
Second Kind" (i.e.incorrectly selecting $H_0$ when $H_1$ is true), for a given rate for ``Errors of the
First Kind" (i.e. rejecting $H_0$ when it is true ).

If, however, one or more of the  hypotheses involves free parameters (`composite hypotheses'),
the Neyman-Pearson lemma does not apply. Nevertheless a form of the likelihood ratio, such 
as the ratio of profile likelihoods, is often used as a method that may well be nearly optimal.

\subsection{$p$-values}
\label{simple p}
For the null hypothesis $H_0$, the expected distribution of our test statistic $t$  is $f_0(t)$. Then for a given  
observed value $t_{obs}$, the $p$-value is the fractional area in the tail of $f_0(t)$ for $t$ greater than or equal to $t_{obs}$. 
For definiteness we consider the single-sided upper tail (assuming that the alternative hypothesis   
yields larger values of  $t_{obs}$), but  lower or 2-sided tails could be appropriate in other cases.


A small $p$-value means that the data are not very consistent with the hypothesis. Apart from the possibility 
that the cause of the discrepancy is new physics, it could be due to an unlikely statistical fluctuation, an incorrect 
implementation of the hypothesis being tested, an inaccurate allowance for detector effects, etc.

As more and more data are acquired, it is possible that a small (and perhaps not physically significant) deviation from 
the tested null hypothesis could result in the $p_0$ becoming small as the data become sensitive to the small deviation. 
For example, a set of particle decays may be expected to follow an exponential 
distribution, but there might be a small background 
characterised by decays at very short times, and which is not allowed for in the analysis. A small amount of data 
might be insensitive to this background, whereas a large amount of data might give a very small $p$-value for a test of
exponential decay, even though the background is fairly insignificant.
 With enough data, we may be able to include physically motivated corrections to our naive $H_0$.
 The possibility of a statistically significant 
but physically unimportant deviation has been mentioned by Cox\cite{Cox_signif}.

It is extremely important to realise that a $p$-value is the probability of observing data like that observed or more extreme, 
assuming the hypothesis is correct. It is {\bf not} the probability of the hypothesis being true, given the data. These are
not the same - see section \ref{PABnePBA}.
    
Many of the negative comments about $p$-values are based on the ease of misinterpreting them. Thus it is possible to 
find 
statements that of all experiments quoting $p$-values below $5\%$, and which thus reject $H_0$, many more than $5\%$ are wrong 
(i.e $H_0$ is actually true). In fact, the expected fraction of these experiments for which $H_0$ is true depends on other 
factors, and could take on any value between zero and unity, without invalidating the $p$-value calculation. 
 
\subsection{$p$-values for two hypotheses}
\begin{figure}[h !]
\begin{center}
\includegraphics[width=.9\textwidth]{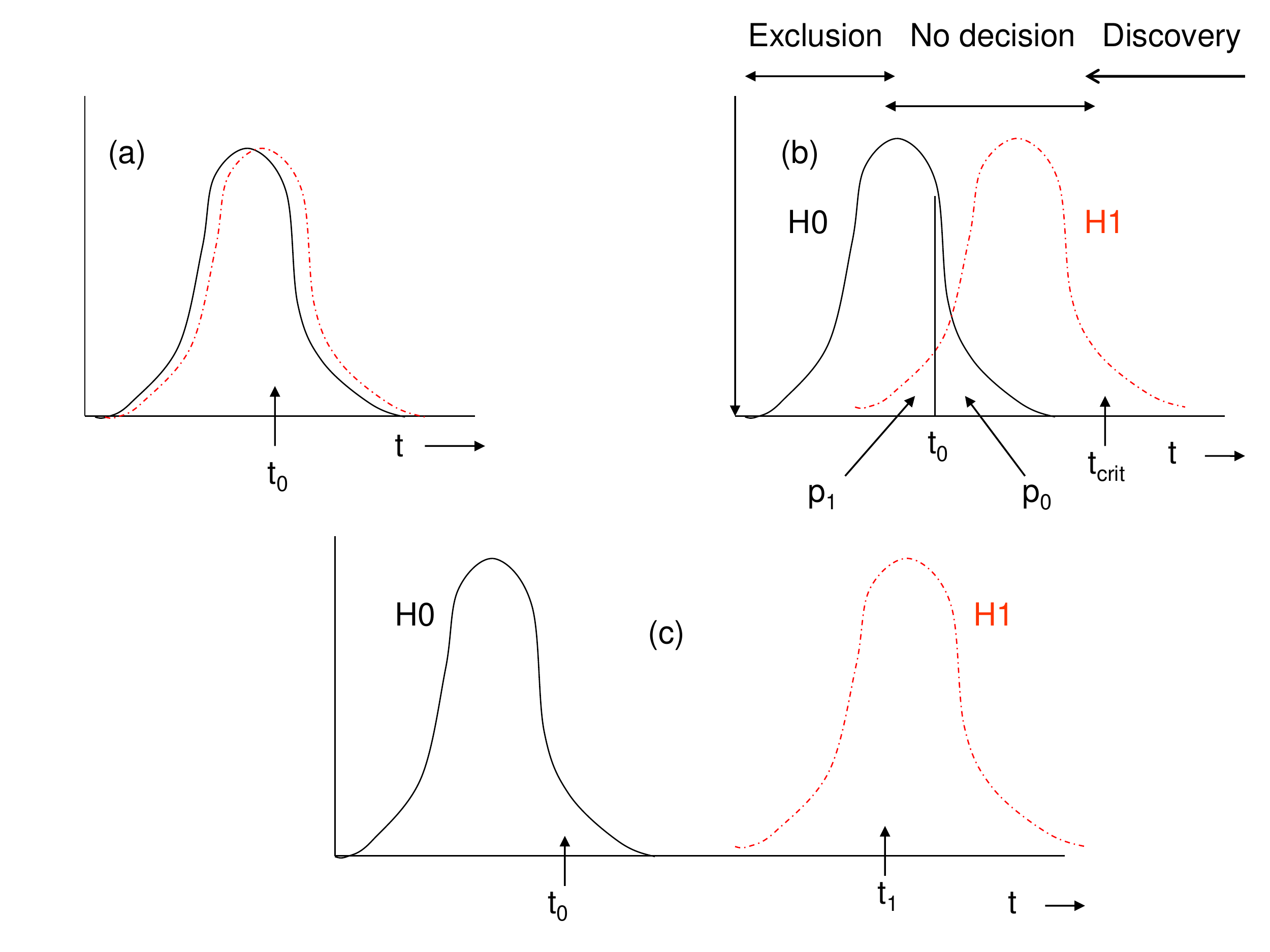}
\caption{
Expected distributions for a statistic $t$ 
for $H_0 =$ background only (solid curves) and for $H_1 =$ background 
plus signal (dashed curves).
In (a), the signal strength is very weak, and it is impossible to choose between $H_0$ and $H_1$. 
As shown in (b), which is for moderate signal strength, $p_0$ is the probability according to $H_0$ of $t$ being 
equal to or larger than the observed $t_0$. 
To claim a discovery, $p_0$  should be smaller than some pre-set level $\alpha$, usually taken to correspond to $5\sigma$; 
$t_{crit}$ is the minimum value of $t$ for this to be so. 
Similarly $p_1$ is the probability according to $H_1$ for $t \leq t_0$. 
The exclusion region corresponds to $t_0$ in the $5\%$ lower tail of $H_1$.
In (b) there is an intermediate ``No decision"  region.
In (c) the signal strength is so large that there is no ambiguity in choosing between the hypotheses.
In order to protect against a downward fluctuation 
in a situation like (a) 
resulting in an exclusion of $H_1$ when the curves are essentially identical, $CL_s$ is defined as
$p_1/(1 - p_0)$ (see Section \ref{CL_s}).
\label{fig:sensitivity}}
\end{center}
\end{figure}

With two hypotheses $H_0$ and $H_1$, we can define a $p$-value for each of them. We adopt the convention that $H_1$ results in 
larger values for the statistic $t$ than does $H_0$. Then $p_0$ is defined as the {\bf upper} tail of $f_0(t)$, the 
probability density function ($pdf$) 
for observing a measured value $t$ when $H_0$ is true. It is conventional to define $p_1$ by the area in the {\bf lower} tail of $f_1(t)$ 
(i.e towards the $H_0$ distribution) -- see Fig. \ref{fig:sensitivity}(b),  which shows the probability densities for obtaining a value 
$t$ of a data statistic, for hypotheses $H_0$ and 
$H_1$. For a 
specific value $t_{obs}$, the $p$-values $p_0$ and $p_1$ correspond to the tail areas above $t_{obs}$ for 
the $H_0$ $pdf$, and 
below   
$t_{obs}$ for $H_1$, respectively.\footnote{If $t$ is a discrete variable, such as a number of events, then `above' 
is replaced by `greater than or equal to', and correspondingly for `below'.} Then $t_{crit}$ is the critical value of $t$ such 
that its $p_0$ value is equal to a preset level $\alpha$ for rejecting the null hypothesis.


%

The $p_1$-value when $t \ = \ t_{crit}$ is denoted by $\beta$, 
and the power of the test is $1 \ - \ \beta$.
The power is the probability that we successfully reject the null hypothesis, assuming that the alternative is true. 
We expect the power to increase as the signal strength in $H_1$ becomes stronger, and the $pdf$s for $H_0$
and $H_1$ become more separated.





 Depending on the separation of the two $pdf$s and on the value of the data statistic $t$, 
several situations are now possible (see Table \ref{table:p_1, p_2}):
\begin{itemize}
\item{$p_1$ is small, but $p_0$ acceptable. Then we accept $H_0$ and reject $H_1$.
i.e.  we exclude the alternative hypothesis.}  
\item{$p_0$ is very small, and $p_1$ acceptable. Then we accept $H_1$ and reject $H_0$. This corresponds to claiming a 
discovery.}
\item{Both $p_0$ and $p_1$ are acceptable. The data are compatible with both hypotheses, and 
we are unable to choose between them.
}
\item{Both $p_0$ and $p_1$ are small. The choice of decision is not obvious, but basically both hypotheses should be 
rejected.}
\end {itemize}  


\fbox{
\parbox {0.9\linewidth}  {
$5\sigma$ DISCOVERY, 95\% EXCLUSION

Searches for new phenomena in Particle Physics typically choose the 
`Standard Model' as the null hypothesis $H_0$, and a specific form
of New Physics as $H_1$. The exclusion level for $H_1$ is usually set at 
5\%, whereas that for rejecting $H_0$ (and perhaps claiming the discovery 
of New Physics) is usually `$5 \sigma$', i.e $p_0 \leq 3\times 10^{-7}$. 

Some (not very convincing) reasons for the stringent criterion for rejecting  $H_0$ include:
\begin{itemize}
\item{The past history of false discovery claims at 3$\sigma$ and 4$\sigma$ levels.}
\item{The possibility that systematic effects have been underestimated.}
\item{The Look Elsewhere Effect (see Section \ref{LEE}).}
\item{Subliminal Bayesian reasoning that the Standard Model is intrinsically 
more likely to be true than some specific speculation about New Physics.}
\item{The embarrassment of having to withdraw a spectacular but incorrect
claim of discovering New Physics.}     
\end{itemize}  
In contrast, incorrect exclusion of New Physics is not regarded as so 
dramatic, and so the weaker criterion of 5\% is used. As Glen Cowan has 
remarked, ``If you are looking for your car keys and are 95\% sure they
are not in the kitchen, it's a good idea to start looking somewhere 
else"\cite{Cowan:keys}.
}}
 
\vspace{0.2in}
 
\begin{figure}[h!]
\begin{center}
\includegraphics[width=0.95\textwidth]{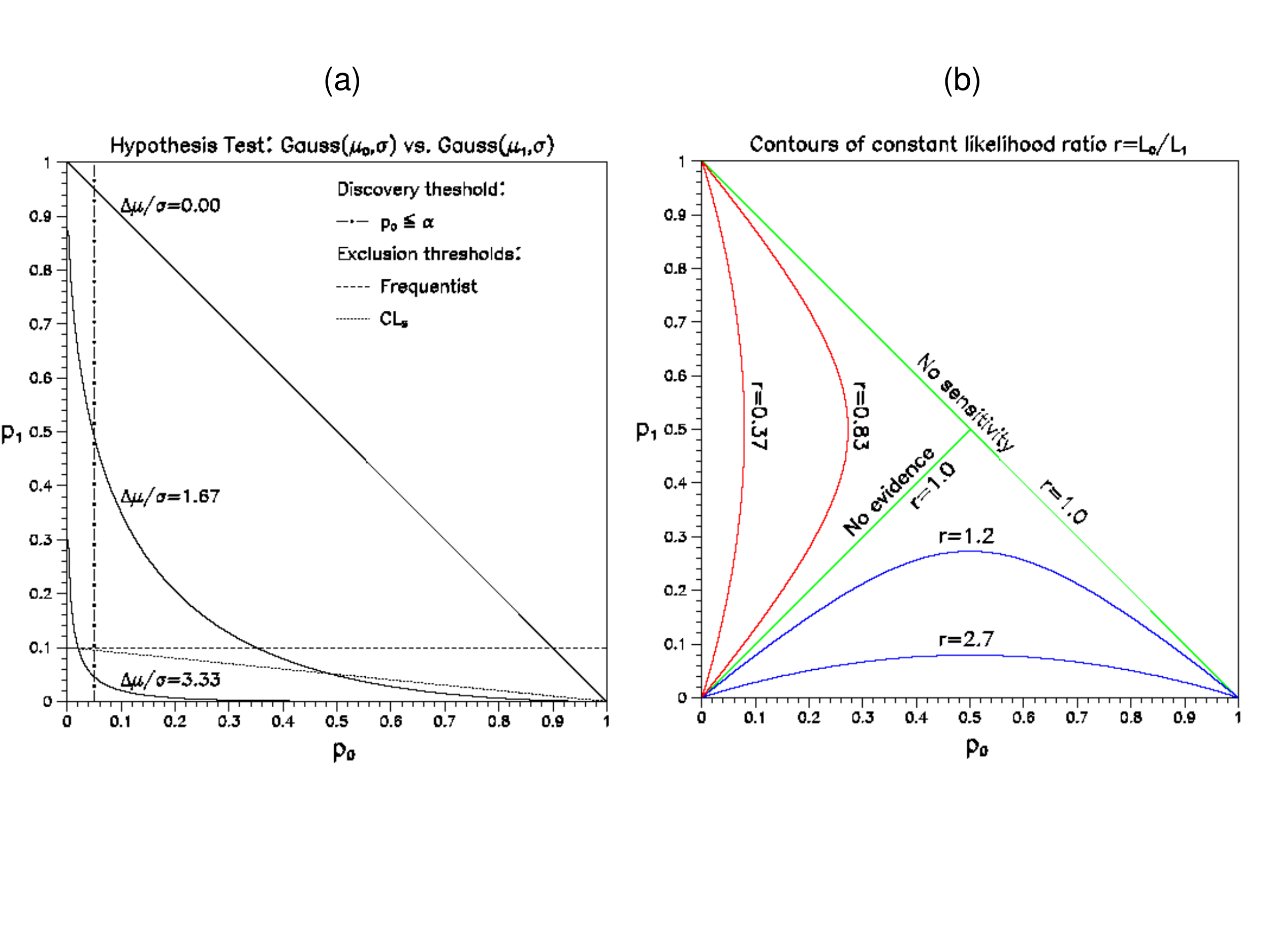}
\caption{Plots of $p_0$ against $p_1$ for comparing a data statistic $t$ with two hypotheses $H_0$ and $H_1$,
whose expected $pdf$s for $t$ are assumed to be two Gaussians of peak separation $\Delta\mu$,
and of equal width $\sigma$ (see fig. \ref{fig:sensitivity}).  
(a) For a  pair of $pdf$s with a given separation, the allowed values of $(p_0, p_1)$
lie on a curve or straight line (shown solid in the diagram). 
As the separation $\Delta\mu$ increases, 
the curves approach the $p_0$ and $p_1$ axes. Rejection of $H_0$ is for $p_0$ less than, say,
$3 \times 10^{-7}$; here it is shown as 0.05 for ease of visualisation. Similarly exclusion of $H_1$
is shown as $p_1 <  0.1$. Thus the $(p_0, p_1)$ square is divided into four regions: 
the largest rectangle is when there is no decision, the long one above the $p_0$-axis is 
for exclusion of $H_1$, the high one beside the $p_1$-axis is for rejection of $H_0$, and
the smallest rectangle is when the data lie between the two $pdf$s. For $\Delta\mu / \sigma = 3.33$,
there are no values of $(p_0, p_1)$ in the ``no decision" region.
In the $CL_s$ procedure, rejection of $H_1$ is when the statistic $t$  is such that $(p_0, p_1)$ 
lies below the diagonal dotted straight line.
(b)  Contours of constant likelihood ratio $r = L_0/L_1$ for Gaussian $pdf$s. The upper right region is inaccessible; 
the diagonal line from (0,1) to (1,0) corresponds to the $pdf$s lying on top of each other i.e. no sensitivity.
The diagonal through the origin is when $t_{obs}$ is mid-way between the two $pdf$s. With larger separation of the
Gaussian $pdf$s and for constant $p_0$ the likelihood ratio increases.  
\label{fig:p0p1}}
\end{center}
\end{figure}

Fig \ref{fig:p0p1}(a) illustrates the $(p_0,p_1)$ plot for defining various decision regions.



\vspace{0.3cm}
\begin{table} [h !]
\caption{Choosing between two hypotheses, based on $p_0$ and $p_1$. }
\begin{center}
\begin{tabular} {|c|c|c|c|c|}
\hline $p_0$ & $p_1$ & Result & If $H_0$ true& If $H_1$ true \\
\hline Very small & O.K. & Discovery  & Error of $1^{st}$ kind & Correct choice\\
\hline O.K. & Small   &   Exclude $H_1$   &  Correct choice &Error of $2^{nd}$ kind \\
\hline O.K.  & O.K. & Make no choice  &    Loss of efficiency & Loss of efficiency\\
\hline Very small  & Small &  ? &     ? & ? \\
\hline
\end{tabular}
\end{center}
\label{table:p_1, p_2}
\end{table}

\subsection{p-values or likelihoods?}
\label{lr}
Rather than calculating $p$-values for the various hypotheses, we could use 
their likelihoods 
$L_0$ and $L_1$. While $p$-values use tail areas beyond the observed statistic, the likelihood is simply the height of the 
$pdf$ at $t_{obs} $. We return to likelihood ratios in Section \ref{Bayesian_HT}.

As mentioned in section \ref{Fr_HT}, the Neyman-Pearson lemma provides the best way of choosing between two simple hypotheses,
but even when one or both hypotheses contain free parameters, 
the likelihood ratio
may well be a suitable statistic 
for summarising the data and for helping choose between the hypotheses. In general, it will be necessary to generate the expected 
distributions of the likelihood ratio according to the hypotheses $H_0$ and $H_1$, in order to make some deduction based on 
the observed likelihood ratio; for composite hypotheses there are of course the complications caused by the nuisance 
parameters.
The decision process may well be based on the $p$-values $p_0$ and $p_1$ for the two hypotheses (see fig. \ref{fig:p0p1}).
In that case, the procedure can be regarded as either a likelihood ratio approach, with the $p$-values simply 
providing a calibration for the value of the likelihood ratio; or as a $p$-value method, with the likelihood ratio
merely being a convenient statistic.

\subsection{Look Elsewhere Effect}
\label{LEE}
If you are playing cards, and in your hand of 13 cards you observe that you
have 4 queens, you might think that that is very unusual. Indeed the probability
of a random set of 13 cards containing 4 queens is 0.0026. However, since you 
decided that `4 queens' was unusual only after you looked at your cards, you might 
have been equally surprised by 4 kings; or 4 jacks; or ace, two, three, four of
the same suit; etc. Taking these into account, the probability of a surprising 
hand of cards similar to what we had is going to be a fair bit larger than 
0.0026. 

A similar effect explains why a seemingly improbable event in our every-day life 
(e.g. a chance meeting with someone we had been thinking about recently) may in fact be
much more likely, if we have not decided at the beginning of the day that this 
specific event would be a real coincidence if it happened.

Often in High Energy Physics, we are looking for some new phenomenon. 
Thus we may be searching for a new particle, whose mass is not 
pre-defined, in a histogram such as that of fig. \ref{fig:Higgs_mass_plot}. When 
we observe an enhancement, we can calculate the $p$-value (the chance of observing 
a statistical fluctuation at least as big as the one in our data, assuming 
that no such particle in fact exists), at the observed 
mass. But this of course underestimates the chance of having a fluctuation anywhere
in our mass distribution, which we might mistakenly ascribe to a new particle. 
We thus need to calculate the probability of observing an effect at least as 
impressive as ours, anywhere in our mass distribution. In Particle Physics, 
this dilution of the significance is known as the Look Elsewhere Effect (LEE).

Similar considerations are relevant for searches in other fields. Thus claims for 
discoveries of gravitational waves would need to calculate the chance of a statistical 
fluctution mimicking the observed effect not only at the observed time, 
frequency and signal structure, but anywhere in the whole dynamic range of these variables 
for which a real signal is possible.

Of course, specifying where exactly `Elsewhere' is is fraught with 
ambiguities. Thus for the above example of searching for a new particle,
fig. \ref{fig:Higgs_mass_plot} is relevant for the possibility of it decaying to
2 photons, but other decay modes could be possible, and hence could be 
relevant to the LEE. Similarly, maybe the particle we are considering
cannot be produced enough at high masses for us to have the chance of 
detecting it, so the whole mass range is not relevant for the LEE. The 
conclusion is that when $p$-values are corrected for the LEE, it is important to
specify exactly what has been taken into account.

\subsection{$CL_s$}
\label{CL_s}
The $CL_s$ method\cite{Read,Junk} was introduced in the LEP experiments at CERN 
in searches for new particles. When evidence for such a particle is not found, the traditional frequentist approach
is to exclude its production if $p_1$ is smaller than some preset level $\gamma$, which in Particle
Physics is typically set at $5\%$. 
However, there is then a $5\%$ probability that $H_1$ could be excluded, even if the experiment was such that the $H_0$
and $H_1 \, pdf$s lay on top of each other i.e. there was no sensitivity to the production of the new phenomenon. 
To protect against this, the decision to exclude $H_1$ is based on $p_1/(1-p_0)$, known as $CL_s$\footnote{This stands for
`confidence level of signal', but it is a poor notation, as $CL_s$ is in fact a ratio of $p$-values, 
which is itself not even a $p$-value, let alone a confidence level.
%
}.
It is thus the ratio of the left hand tails 
of the $pdf$s for $H_1$ and $H_0$. Fig \ref{fig:p0p1}(a) shows a $(p_0,p_1)$ region for which $H_1$ is excluded by $CL_s$. 
The fact that it is clearly smaller than for the standard frequentist exclusion region is the price one has to pay 
for the protection it provides against excluding $H_1$ when an experiment has no sensitivity to it. 
We regard it as conservative frequentist.

It is interesting that the $CL_s$ exclusion line in fig. \ref{fig:p0p1}(a) for the case of two Gaussians is identical 
to that obtained by a Bayesian procedure for determining the upper limit on $\mu_1$ when the latter is restricted 
to positive values, and with a uniform prior for $\mu_1$. In a similar manner, the standard frequentist procedure agrees 
with the Bayesian upper limit when  the restriction of $\mu_1$ being positive is removed.     

In principle, similar protection against discovery claims when the experiment has no sensitivity could be employed,
but it is deemed not to be necessary because of the different levels used for discovery and for exclusion of 
$H_1$ (typically $3 \times 10^{-7}$ and 0.05 respectively).

\subsection{When neither $H_0$ or $H_1$ is true}
It may well be that neither $H_0$ nor $H_1$ is true. With no more information available,
it is of course impossible to say what we expect for the distribution of our test statistic $t$. 
On the plot of fig. \ref{fig:p0p1}(a), our data may fall in the small rectangle next to the origin. 

It is certainly not true that a  small value for $p_0$ necessarily implies that $H_1$ is correct, although for small enough
$p_0$, ruling out $H_0$ is a possibility.


\section{BAYESIAN METHODS}
\label{Bayesian_HT}
The Bayesian approach is more naturally suited to making statements about what we believe about two (or more) hypotheses
in the light of our data. This contrasts with Goodness of Fit, which involves considering other possible data outcomes,
but focusses on just one hypothesis.

All Bayesian methods involve the likelihood function, possibly modified to take into account  nuisance parameters.
For Hypothesis Testing, some form of a ratio of (modified) likelihoods is usually involved.
For simple hypotheses, this is just 
$L_0(H_0)/L_1(H_1)$, where $L_i(H_i) = p(x|H_i)$, the probability (density) for observing data $x$ for the hypothesis $H_i$. 
The issue is going to be how nuisance parameters\footnote{For the purpose of model comparison, any parameters
are considered as nuisance parameters, even if they are physically meaningful. e.g. the parameters of a straight line fit, 
the mass of the Higgs boson, etc.} $\nu$ are dealt with for non-simple hypotheses. For the likelihood approach 
(as opposed to the Bayesian one, which also requires priors), it is usual to profile them i.e.
the profile likelihood is $L_i(H_i|\nu_{best})$,
where $\nu _{best}$ is the set of parameters which maximise $L$. 
In Particle Physics, the profile likelihood approach 
is a popular method for incorporating systematics in parameter determination problems.

The complications of applying Bayesian methods to model selection in practice are due to the choices for appropriate priors. 
This is particularly so for those parameters which occur in the alternative hypothesis $H_1$ but not in the null $H_0$.

Loredo\cite{Loredo} and Trotta\cite{Trotta} have provided reviews of the application of Bayesian techniques 
in Astrophysics and Cosmology, where their use is more common than in Particle Physics.

\subsection{Bayesian posterior probabilities}
When there are no nuisance parameters involved, the
ratio of the posterior probabilities for $H_i$ is $p_{post}(H_0|x) / p_{post}(H_1|x)$, where 
\begin{equation}
p_{post}(H_i|x) = L_i(H_i)\ \pi_i ,
\end{equation}
and $\pi_i$ is the assigned prior probability for hypothesis $i$. For example, the hypothesis of there being a Higgs boson
of mass 110 GeV might well be assigned a small prior, in view of the exclusion limits from LEP.

With nuisance parameters, the posterior probabilities become
\begin{equation}
p_{post}(H_i|x) = \int L_i(H_i|\nu)\ \pi_i(\nu)\ \pi_i\ d\nu 
\end{equation}
where $\pi_i$ is the prior probability for hypothesis $i$ and $\pi_i(\nu)$ is the joint prior for its nuisance parameters. i.e. 
we now have {\bf integrated} over
the 
nuisance parameters. This contrasts with the likelihood method, where {\bf maximisation} with respect to them is more usual. 
Even with $\pi_i(\nu)$ being a constant, integration and maximisation can select different regions of parameter space. 
An example of this would be a likelihood function that has a large narrow spike at small $\nu$, 
and a broad but lower enhancement at large $\nu$.  

In relation to all Bayesian methods, it is to be emphasised that the choice of a constant prior, especially for 
multi-dimensional $\nu$, is by no means obvious (compare Section \ref{priors}). 
Very often, there are several possible choices of variable 
for the nuisance parameters, with none of them being obviously more natural or appropriate that the others. Thus a point in
2-dimensional space could be written as Cartesian $(x,y)$ or polar $(r,\theta)$; constant priors in the two sets of variables 
are different. Similarly in fitting data by a straight line $y = a + b\, x$, using a seemingly innocuous flat prior for 
$b = \tan\theta$ results in angles $\theta$ in the range $0^{\circ}$ to $89^{\circ}$ have 
the same prior probability as those in the range $89^{\circ}$ to $89.5^{\circ}$.   

It should be realised that the results for Hypothesis Testing are more sensitive to the choice of prior than in parameter 
determination. Thus in parameter determination, sometimes a prior is used which is constant over a wide range of $\nu$, 
and zero outside it. The resulting range for the parameter, as deduced from its posterior, may well be insensitive 
to the range used, provided it includes the region where the likelihood $L(\nu)$ is significant. 
For comparing hypotheses, however, there can be parameters which occur in one hypothesis but not the other. 
(An example of this is where $H_1$ corresponds to smooth background plus a peak, while $H_0$ is just smooth background.)
The widths of such priors affect their normalisation, and hence also the Bayes factor (see next Section) directly. 

On the other hand, in searches for a new particle of unknown mass, the Bayesian
probability for the particle existing will depend on the range of the prior used
for the particle's mass - the wider the range, the lower the normalisation and 
hence the lower the probability\footnote{This is an example of Occam's Razor,
whereby a simpler hypothesis may be favoured over a more complex one.}. 
At least qualitatively, this resembles the effect of the LEE in the frequentist
approach, where the significance of a peak in a mass spectrum is diluted if the search
extends over a wide mass range (see section \ref{LEE}).

\subsection{Bayes factor}
\label{Bayes_factor}
For each hypothesis we define $R_i = p_{post}/\pi$, where $p_{post}$ and $\pi$ are respectively 
the posterior and prior probabilities for hypothesis $i$. Thus $R$ is just the ratio of posterior and prior probabilities. 
Then the Bayes factor for the two hypotheses $H_0$ and $H_1$ is $B_{01} = R_0/R_1$. If the two hypotheses are both simple,
then this is just the likelihood ratio. If either is composite, the relevant integrals are required for $p_{post}$.
A small value of $B_{01}$  favours $H_1$.


Demortier\cite{luc_mbf} has drawn attention to the fact that it can be useful to calculate the {\bf minimum} Bayes 
factor\cite{ELS}. This is defined as above, but with the extra nuisance parameters of $H_1$ set at values that 
minimise $B_{01}$, i.e. they are as favourable as possible for $H_1$. If even this value of $B_{01}$ suggests that $H_1$
is not to be preferred, then it is a waste of time to investigate further since any choice of priors 
for the extra parameters cannot make $B_{01}$ smaller.  

%

\subsection{Other Bayesian methods}
The Bayesian approach can be used in conjunction with Decision Theory,
in order to provide a procedure for choosing between two hypotheses. In addition to any priors,  a
cost function has to be defined, which assigns a numerical `cost' for
each combination of the true hypothesis ($H_0$ or $H_1$), and the possible
decisions - see Table \ref{Table:Cost}. The decision procedure is 
designed to minimise the expected cost, as determined by the cost 
function and the expected distribution of posterior probabilities for $H_0$
and $H_1$. 

\begin{table} 
\caption{Cost function. Typically the cost $A$ assigned to a false discovery claim
would be larger than $B$, the cost for a failure to discover. There is zero cost for 
making a correct decision.}
\begin{center}
\begin{tabular} {|p{2cm}|p{5cm}|p{5cm}|}
\hline  & $H_0$ true &  $H_1$ true \\
\hline Accept $H_0$  & Correct choice. Cost =0  &      Failure to discover. Cost = $B$ \\
\hline Reject $H_0$  & False discovery claim. Cost = $A$ &   Correct choice. Cost = 0\\
\hline
\end{tabular}
\end{center}
\label{Table:Cost}
\end{table}

Because of the problems of assigning realistic costs, and the use of 
priors in determining the posteriors for the hypotheses, there is
little or no usage of this approach in Particle Physics searches for New Physics.

Other Bayesian methods such as AIC, BIC,.... (Akaike Information Criterion,
Bayesian Information Criterion,.....) aim to provide approximations to the Bayes
factor, but which are easier to calculate. Given the powerful computational  
facilities available nowadays, these methods are likely to decrease in general 
usage. Again there is little or no experience of using them in Particle 
Physics applications.

\subsection{Why $p$ is not equal to the likelihood ratio}
\label{pnotB}
There is sometimes discussion of why a likelihood ratio approach (or the Bayes factor,
if there are nuisance parameters) can give a very different numerical
answer to a $p$-value calculation. A reason some agreement might be expected is that they are both addressing the question of
whether there is evidence in the data for new physics. 

In fact they measure very different things. Thus $p_0$ simply measures the consistency with the null hypothesis, 
without any regard to the degree of agreement with the alternative, while the likelihood ratio takes the alternative into 
account.
There is thus no 
reason to expect them to bear any particular relationship to each other. This can be illustrated by contours of 
constant values of the likelihood ratio $r=L_0/L_1$  on a $p_0$ versus $p_1$ plot (see fig. \ref{fig:p0p1}(b)). The figure is 
constructed by assuming that the 
$pdf$s for the two hypotheses $H_0$ and $H_1$ are given by Gaussian distributions, both of unit width.
 Then at constant $p_0$, it  is 
seen that the likelihood ratio can take a range of values, corresponding to the Gaussians having different separations. Thus with the  
Gaussian for $H_0$'s $pdf$ centred at zero, a measured value of 5.0 yields a $p_0$-value of $3 \times 10^{-7}$,
regardless 
of the position of the $H_1$ Gaussian. Such a small $p$-value is usually taken as sufficient to reject $H_0$. 
As the centre of the $H_1$ Gaussian starts at $\mu_1 = 0$, the two Gaussian $pdf$s are identical, and $r=1$.
With increasing $\mu_1$, $p_0$ of course remains constant, but $r$
at first decreases to a minimum when $\mu_1 = 5$, but then increases through unity when $\mu_1 = 10$
(i.e. the data is midway between the $pdf$ peaks), and then keeps on rising with further increases of the 
 separation of the $pdf$s.
 At that stage, the data are more in agreement with $H_0$ than with $H_1$, 
despite the small value of $p_0$.

\section{CONCLUSION}

\begin{table}[h!]
\caption{Comparison of Bayes and Frequentist approaches}
\begin{center}
\begin{tabular} {|p{3cm}|p{5cm}|p{5cm}|}
\hline
                &    Bayes   &                    Frequentist \\
\cline{2-3}
Basis of method   & Bayes Theorem $\rightarrow$ posterior probability distribution     &    Uses $pdf$ for data, for fixed parameter values \\
\hline
Meaning of probability & Degree of  belief & Frequentist definition\\
\hline
Probability for parameters?  &    Yes    &  No, no, no  \\
\hline
Needs prior?    &  Yes&   No\\
\hline
Choice of interval? & Central, upper limit, shortest,...    &   Choice of ordering rule \\
\hline
Data used   &   Only the data you have      &    Also other possible data  \\
\hline
Needs ensemble of possible experiments?   &   No   &    Yes (but often not explicit)  \\
\hline
Obeys the Likelihood Principle?     &   Yes     &   No    \\
\hline
Unphysical/empty ranges possible?     &    Excluded by prior       &     Can occur  \\
\hline
Final statement    &   Posterior prob dist    & Param values for which data is likely    \\
\hline
Do param ranges cover?    &   Regarded as unimportant   &    Built in   \\
\hline
Include systematics   &   Integrate over prior     &   Extend dimensionality of frequentist construction   \\
\hline
\end{tabular}
\end{center}
\label{comparison}
\end{table}                
   
We  have seen how Bayesians and Frequentists differ fundamentally in the way they
consider probability. This then affects the way they approach the topics of parameter
determination, and of choosing between two hypotheses. Table \ref{comparison}  provides a summary 
of the differences between the two approaches.

A cynic's view of the two techniques is provided by the quotation:


``Bayesians address the question everyone is interested in by using assumptions 
no-one believes, while
Frequentists use impeccable logic to deal with an issue of no interest to anyone."

However, it is not necessary to be so negative, and for physics analyses at the CERN's
 LHC, the aim is, at least for determining parameters and setting upper limits in 
searches for various new phenomena, to use both approaches; similar answers would 
strengthen confidence in the results, while differences suggest the need to 
understand them in  terms of the somewhat different questions that the two approaches 
are asking. 

It thus seems that the old war between the two methodologies is subsiding, and that they
 can hopefully live together in fruitful cooperation.

\vspace{0.1in}
ACKNOWLEGEMENTS

I would like to thank Bob Cousins, David van Dyk, Luc Demortier and Roberto Trotta for their advice on various sections of this article.



%

%
\end{document}